\documentclass[prd,twocolumn,nopacs,floatfix,amsmath,nofootinbib,amssymb,floatfix]{revtex4}
\usepackage{graphicx,color,dcolumn,booktabs,bm}
\usepackage{longtable,lscape}
\usepackage{txfonts}
\usepackage{overpic}
\usepackage{amssymb}
\usepackage{amsmath}
\usepackage{indentfirst}
\usepackage{feynmf}   
\usepackage{slashed}  
\usepackage{cases}
\usepackage{color}
\usepackage{multirow}
\usepackage{epstopdf}
\usepackage{longtable}
\usepackage{tikz}
\usepackage{graphicx,color,dcolumn,booktabs,bm}
\usepackage[colorlinks,
            citecolor=blue,
            anchorcolor=red,
            menucolor=red,
            linkcolor=red,
            filecolor=red,
            runcolor=red,
            urlcolor=blue,
            frenchlinks=red]{hyperref}

\graphicspath{{Figures/}} %

\allowdisplaybreaks

\begin{document}

\title{Exploring pion emission properties of (strange) hidden-charm molecular pentaquarks in a chiral quark model}
\author{Li-Cheng Sheng$^{1,3}$}
\author{Yu-Jie Tang$^{1}$}
\author{Yu-Xin Wan$^{1}$}
\author{Rui Chen$^{1,2}$\footnote{Corresponding author}}\email{chenrui@hunnu.edu.cn}
\author{Dian-Yong Chen$^{3}$}\email{chendy@seu.edu.cn}

\affiliation{
$^1$Key Laboratory of Low-Dimensional Quantum Structures and Quantum Control of Ministry of Education, Department of Physics and Synergetic Innovation Center for Quantum Effects and Applications, Hunan Normal University, Changsha 410081, China\\
$^2$Hunan Research Center of the Basic Discipline for Quantum Effects and Quantum Technologies, Hunan Normal University, Changsha 410081, China\\
$^3$ School of Physics, Southeast University, Nanjing 210094, People’s Republic of China}
\date{\today}
\begin{abstract}
The internal structure of the exotic $P_c$ and $P_{cs}$ pentaquarks remains an open question. To address this, we demonstrate that pion emission serves as a sensitive probe by calculating its properties within a molecular scenario using the chiral quark model and coupled-channel effects. Our results reveal a strong dependence of the decay widths on the internal structure and spatial wave functions. We therefore expect this study to stimulate experimental measurements of these decays, which are crucial for determining the nature of these states and guiding the search for further molecular partners.
\end{abstract}
\pacs{***}
\maketitle

\section{introduction}

Over the past twenty years, experiments have observed numerous exotic hadronic structures that are distinct from conventional mesons (composed of a quark-antiquark pair) and baryons (composed of three quarks). Studying these states cannot only help to establish and refine the hadron spectrum but also provide a powerful approach for deepening our understanding of complex non-perturbative phenomena in quantum chromodynamics (QCD).

A prime example of these exotic structures is the family of hidden-charm pentaquarks. In 2015, the LHCb Collaboration reported the first observations of two hidden-charm structures, labeled as $P_c(4380)$ and $P_c(4450)$, in $\Lambda_b^0\to J/\psi pK^-$ decay process \cite{LHCb:2015yax}. Theoretically, the discovery of $P_c(4380)$ and $P_c(4450)$ stimulated vigorous discussions concerning their internal structure and underlying production mechanisms. Various interpretations were proposed, including molecular hadronic states, compact pentaquarks, and non-resonant explanations\cite{Chen:2015loa,Chen:2015moa,Karliner:2015ina,Maiani:2015vwa,Lebed:2015tna,Wang:2015ava}. Given that the masses of $P_c(4380)$ and $P_c(4450)$ lie close to the thresholds of charmed baryon and anti-charmed meson pairs, the molecular state hypothesis emerged as a popular explanation. In fact, theorists had predicted the existence of such hidden-charm molecular pentaquarks prior to their experimental observation\cite{Wu:2010jy,Yang:2011wz,Wang:2011rga,Wu:2012md,Li:2014gra}. Detailed reviews of the theoretical progress on these initial $P_c$ states can be found in Refs. \cite{Chen:2016qju,Dong:2017gaw,Guo:2017jvc}. This theoretical landscape, however, was reshaped by subsequent experimental data.

With the accumulation of more data, the LHCb Collaboration performed a new analysis of the same decay channel in 2019. This analysis not only revealed a new narrow structure, $P_c(4312)$, but also resolved the previously observed $P_c(4450)$ into two distinct substructures, labeled as $P_c(4440)$ and $P_c(4457)$ \cite{LHCb:2019kea}. The discovery of these fine structures provided stronger support for interpreting hidden-charm pentaquarks as meson-baryon molecular states. Although the molecular picture gained significant traction following the 2019 results, it is not the only viable interpretation. Several studies continue to argue that these $P_c$ states are compact pentaquarks \cite{Ali:2019npk,Mutuk:2019snd,Wang:2019got,Cheng:2019obk,Weng:2019ynv,Pimikov:2019dyr}.

Beyond the fundamental question of molecular versus compact structure, a further debate exists regarding the specific quantum numbers of these states within the molecular framework. For example, in our previous work \cite{Chen:2019asm}, we investigated the interactions between a charmed baryon and an anti-charmed meson using the one-boson-exchange (OBE) model, including coupled-channel effects. We found that $P_c(4312)$ can be identified as the $\Sigma_c\bar{D}$ molecule with $I(J^P)=1/2(1/2^-)$, while $P_c(4440)$ and $P_c(4457)$ can be identified as the $\Sigma_c\bar{D}^*$ molecules with $1/2(1/2^-)$ and $1/2(3/2^-)$, respectively. In Ref. \cite{Chen:2019bip}, the authors proposed that $P_c(4312)$ can be well interpreted as the $\Sigma_c^{++}D^-$ bound state with $J^P=1/2^-$, while $P_c(4440)$ and $P_c(4457)$ could be interpreted as the $\Sigma_c^+\bar{D}^0$ bound state with $1/2^-$, the $\Sigma_c^{*++}D^-$ and $\Sigma_c^+\bar{D}^{*0}$ bound states with $3/2^-$, or the $\Sigma_c^{*+}\bar{D}^{*0}$ bound state with $5/2^-$, based on their previous QCD sum rule studies. 

This pattern of discovery followed by theoretical ambiguity was repeated with the observation of a strange hidden-charm pentaquark. In 2020, the LHCb Collaboration reported evidence for the strange hidden-charm pentaquark $P_{cs}(4459)$ in $\Xi_b^-\to J/\psi\Lambda K^-$ with a statistical significance of 3.1 $\sigma$ \cite{LHCb:2020jpq}. Theoretically, Theoretically, various explanations for $P_{cs}(4459)$ have been proposed, including molecular states, compact pentaquarks, and triangle singularity mechanisms \cite{Zou:2021sha,Karliner:2021xnq,Peng:2020hql,Zhu:2021lhd,Hu:2021nvs,Xiao:2021rgp,Deng:2022vkv,Wang:2020eep,Ferretti:2021zis,Azizi:2021utt,Shen:2020gpw}. Because $P_{cs}(4459)$ lies just below the $\Xi_c\bar{D}^*$ threshold, the molecular scenario has attracted considerable attention. Nevertheless, there is ongoing discussion about the internal structure of $P_{cs}(4459)$ as a hadronic molecule \cite{Peng:2020hql,Li:2021ryu,Wu:2021caw,Lu:2021irg,Yang:2021pio}. For instance, since the spin-parity of $P_{cs}(4459)$ has not been determined experimentally, a key theoretical question is whether it can be assigned as a $\Xi_c\bar{D}^{(*)}$ molecule with $I(J^P)=0(1/2^-)$ or $0(3/2^-)$ \cite{Lu:2021irg,Du:2021bgb,Ozdem:2021ugy,Gao:2021hmv}.  In Ref. \cite{Chen:2022onm}, we performed a coupled-channel analysis of $\Xi_c\bar{D}^*/\Xi_c^*\bar{D}/\Xi_c^{\prime}\bar{D}^*/\Xi_c^*\bar{D}^*$ using the OBE model. We found that $P_{cs}(4459)$ can be described as a bound state in this coupled-channel system with $I(J^P)=0(3/2^-)$, where the dominant components are $\Xi_c\bar{D}^*$ and $\Xi_c^*\bar{D}$. Beside the $\Xi_c\bar{D}^*$ molecular explanation, A. Feijoo \textit{et al.} proposed that $P_{cs}(4459)$ is associated to the state coupled mostly to $\Xi_c^{\prime}\bar{D}$ \cite{Feijoo:2022rxf}, after studied the interaction in coupled channels of pseudoscalar mesons and vector mesons with baryons of $J^P=1/2^+$ and $3/2^+$.

\begin{figure}[!htbp]
    \centering
    \includegraphics[width=0.9\linewidth]{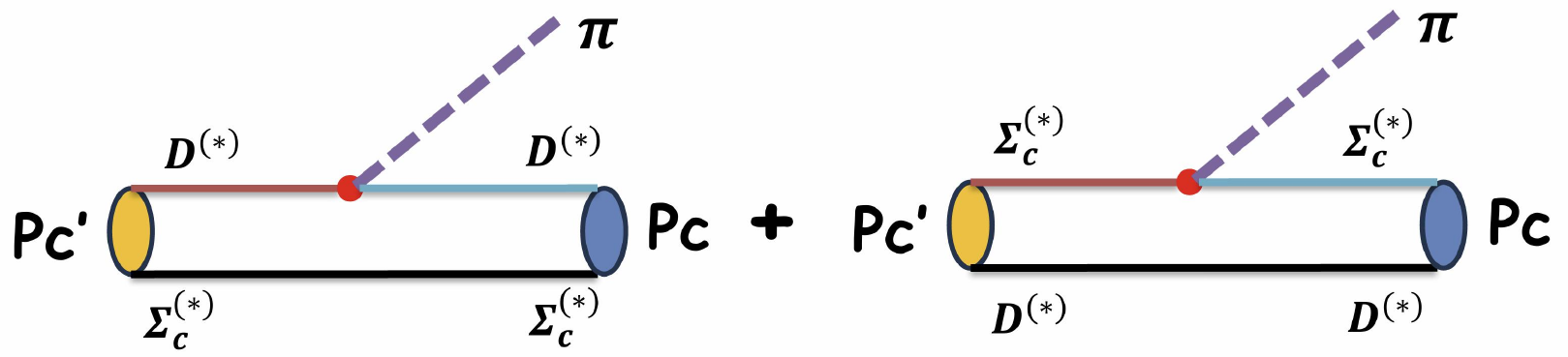}
    \caption{The $\pi$ emission process between the $P_c$ molecular states.}
    \label{decay}
\end{figure}

The cumulative evidence from the $P_c$ and $P_{cs}$ states thus presents a clear challenge: while abundant theoretical models exist, a definitive method to pinpoint their internal structure is lacking. Therefore, a more comprehensive and systematic investigation is required. As is well known, decay properties are highly sensitive to the wave functions of the interacting particles. Consequently, we propose that exploring pion $\pi$ emission properties provides a crucial and sensitive probe. Since $\pi$ emission involves spin-related interactions, our results can help identify molecular configurations, especially their spin-parity quantum numbers. In the meson-baryon molecular scenario, the 
$\pi$ emission occurs at the constituent hadrons, as illustrated in Figure \ref{decay}. {In Ref. \cite{Chen:2016byt}, the authors investigated similar processes, namely the pionic transition from $Y(4260)$ to $Z_c(3900)$.}

To this end, in this work we adopt the chiral quark model to derive the decay amplitudes for the $\pi$ emission processes between possible $P_c/P_{cs}$ molecules. In addition to focusing on the reported $P_c/P_{cs}$ states as molecules, we also calculate the $\pi$ emission decay widths for transitions between the reported states and other possible (strange) hidden-charm molecular pentaquarks predicted in our previous works \cite{Chen:2019asm,Chen:2022onm}. Through these efforts, we aim to provide new methods for searching for possible hidden-charm molecular partners.

This paper is organized as follow. After introduction, the detailed deduction of the $\pi$ emission interactions between possible $P_c/P_{cs}$ states molecules systems is given in Sec. \ref{sec2}. In Sec. \ref{sec3}, we present the numerical results. The paper ends with a summary in Sec. \ref{sec4}.

\section{Framework}\label{sec2}

In this work, we adopt the chiral quark model to deduce the $\pi$ emission interactions. The corresponding effective Lagrangians can be constructed as follows \cite{Manohar:1983md,Wang:2017kfr,Zhong:2008kd,Xiao:2017udy,Ni:2023lvx,Zhou:2023wrf,Zhong:2024mnt,Zhong:2007gp}:
\begin{align}
\mathcal{L} &= \frac{\delta}{\sqrt{2} f_{p}} \bar{\psi}_j \gamma_\mu \gamma_5 \psi_j \vec{I} \cdot \vec{\partial}^\mu {\phi}_p.
\end{align}
Here, $I$ is an isospin operator. $\delta$ and $f_{p}$ denote the coupling strength for the quark-pseudoscalar-meson interaction and the pseudoscalar meson decay constant, respectively. ${\phi}_p$ and $\psi_j$ represent the pseudoscalar meson field and the $j$th light quark field, respectively. In a Pauli-Dirac representation, the quark field can be approximated as
\begin{align}
\psi_j &\simeq \begin{pmatrix} 
1 \\ 
\frac{\boldsymbol{\sigma}_j \cdot \boldsymbol{P}_i}{E_i + M_i} + \frac{\boldsymbol{\sigma}_j \cdot \boldsymbol{P}_j}{2m_j}
\end{pmatrix},
\end{align}
where $\boldsymbol{P}_i$ is the momentum of the initial molecular state. The quantities $\boldsymbol{P}_j$, $\boldsymbol{\sigma}_j$, and $m_j$ are the internal momentum, spin operator, and mass of the $j$th quark within the molecular state, respectively.

Through a non-relativistic reduction, the non-relativistic form of $\mathcal{L}$ up to order $1/m$ can be obtained:
\begin{align}
\mathcal{L} &= g \sum_{j} \left( \mathcal{G} \boldsymbol{\sigma}_j \cdot \boldsymbol{q} + h \boldsymbol{\sigma}_j \cdot \boldsymbol{p}_j \right) F(\boldsymbol{q}^2) I_j \varphi_m.\label{H}
\end{align}
In the Eq.(\ref{H}), $\boldsymbol{\sigma}_j$ and $\boldsymbol{p}_j$ are the spin operator and momentum operator of the $j$th light quark in the molecular state. $\varphi_m=e^{-i\boldsymbol{q}\cdot\boldsymbol{r}_{j}}$ is the plane wave function of the emitted light meson. The constants are defined as $g= i \delta \sqrt{(E_i + M_i)(E_f + M_f)}/(\sqrt{2} f_\mathcal{M})$, $\mathcal{G}= -\left( \frac{\omega_m}{E_f + M_f} + 1 \right)$, and $h=\omega_m/2\mu_q$, where $(E_i,M_i)$ and $(E_f,M_f)$ are the energy and mass of the initial and final molecular states, respectively. $\omega_m$ and $\boldsymbol{q}$ are the energy and three momentum of the emitted meson, respectively. The reduced mass $\mu_q$ is expressed as $\mu_q=m_jm_{j^\prime}/(m_j+m_{j^\prime})$, where $m_j$ and $m_{j^\prime}$ are the mass of $j$-th quark in the initial and final molecular states. Following Ref.~ \cite{Zhong:2008kd,Zhong:2007gp}, the isospin operator $I_j$ is given by 
\begin{align}
I_j = 
\begin{cases} 
a_j^\dagger (u) a_j (d) & \text{for } \pi^-, \\
a_j^\dagger (d) a_j (u) & \text{for } \pi^+, \\
\dfrac{1}{\sqrt{2}} \big[ a_j^\dagger (u) a_j (u) - a_j^\dagger (d) a_j (d) \big] & \text{for } \pi^0, \\
\end{cases}
\end{align}
where $a_j^\dagger (u,d)$ and $a_j (u,d)$ are the creation and annihilation operators for $u$ and $d$ quarks. To suppress unphysical high-momentum contributions, a form factor $F(\boldsymbol{q}^2)$ is introduced, expressed as $F(\boldsymbol{q}^2)=\sqrt{\Lambda^2/(\Lambda^2+\boldsymbol{q^2})}$.

The model parameters used in this work are listed in Table~\ref{parameters}. As in Refs.~ \cite{Zhong:2008kd,Zhong:2007gp}, the quark-meson coupling strength is set to $\delta=0.557$. The cut-off parameter $\Lambda$ is taken as $\Lambda=0.66$ GeV, following the discussions in Refs.~ \cite{Zhong:2024mnt,Ni:2023lvx,Yu:1995ag,Vijande:2004he,Valcarce:2008dr}.
\begin{table}[!htbp]
    \centering
    \renewcommand\tabcolsep{0.05cm}
    \renewcommand{\arraystretch}{1.5}
    \caption{The parameters adopted in this work.}
    \label{parameters}
    \begin{tabular}{ccccccc}
    \toprule[1pt]
    \toprule[1pt]
       $m_{u}$(GeV) &$m_{d}$(GeV) &$m_{s}$(GeV) &$m_{c}$(GeV) &$\delta$  &$\Lambda$(GeV) &$f_{\pi}$(GeV)  \\ \hline
        $0.45$ &$0.45$ &$0.55$  &$1.68$ &$0.557$  &$0.66$ &$0.093$ \\ \bottomrule[1pt]\bottomrule[1pt]
    \end{tabular}
\end{table}

For the $\pi$ meson emission between possible $P_{c(s)}$ molecules, the strong decay amplitude is given by
\begin{align}
\mathcal{M}[\mathcal{B} \to \mathcal{B}^\prime + \pi ] = \langle \mathcal{B}^\prime | \mathcal{H}^{NR} | \mathcal{B} \rangle,
\end{align}
where $\mathcal{B}$ and $\mathcal{B}^\prime$ are the initial and final molecular states. The corresponding partial decay width is calculated as
\begin{align}
\Gamma = \frac{1}{8\pi} \frac{|\boldsymbol{q}|}{M_i^2} \frac{1}{2J_i + 1} \sum_{J_{f_z}} \left| \mathcal{M}_{J_{i_z} J_{f_z}} \right|^2.
\label{decay_width}
\end{align}
Here, $J_i$ is the total angular momentum quantum number of the initial molecular state, and $J_{i_z}$ and $J_{f_z}$ are the third components of the total momenta of the initial and final molecular states, respectively.

\begin{figure}[h]
    \centering
   \includegraphics[width=0.7\linewidth]{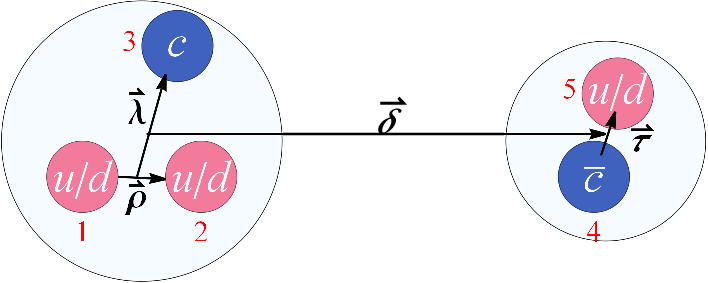}
    \caption{The Jacobi coordinates of the $\Sigma_c\bar{D}^{(*)}$ molecular states.}
    \label{jacobi}
\end{figure}

In Fig.~\ref{jacobi}, Jacobi coordinates $\boldsymbol{R}=\boldsymbol{\rho}, \boldsymbol{\lambda}, \boldsymbol{\tau}, \boldsymbol{\delta}$ are used to describe the spatial wave function of the hadronic molecular states. The total spatial wave function for the initial and final states is then expressed as a product: $\phi(\boldsymbol{\rho})\phi(\boldsymbol{\lambda})\phi(\boldsymbol{\tau})\phi(\boldsymbol{\delta})$. These Jacobi coordinates are related to the quark coordinates $\bm{r}_j$ by
\begin{align}
\boldsymbol{\rho}&=\boldsymbol{r}_2-\boldsymbol{r}_1, \nonumber\\
\boldsymbol{\lambda}&=\boldsymbol{r}_3-\frac{m_1 \boldsymbol{r}_1+m_2 \boldsymbol{r}_2}{m_1+m_2}, \nonumber\\
\boldsymbol{\tau}&=\boldsymbol{r}_5-\boldsymbol{r}_4, \nonumber\\
\boldsymbol{\delta}&=\frac{m_4 \boldsymbol{r}_4+m_5 \boldsymbol{r}_5}{m_4+m_5}-\frac{m_1 \boldsymbol{r}_1+m_2 \boldsymbol{r}_2+m_3 \boldsymbol{r}_3}{m_1+m_2+m_3}.
\end{align}

The spatial wave function between the constituent hadrons, $\phi(\boldsymbol{\delta})$, is obtained by solving the Schrödinger equation. We employ the simple harmonic oscillator (SHO) wave function to describe the internal spatial wave functions of the mesons and baryons \cite{Li:2008xy,Liu:2011yp,Lai:2024jfe,Zhang:2025ame,Wang:2022nqs}:
\begin{align}
\phi_{n,l,m}(\beta, \mathbf{r}) =& \sqrt{\frac{2n!}{\Gamma\left(n + l + \frac{3}{2}\right)}} L_n^{l + \frac{1}{2}} \left( \beta^2 r^2 \right) \beta^{l + \frac{3}{2}} \mathrm{e}^{-\frac{\beta^2 r^2}{2}} r^l Y_{lm}(\Omega).
\label{SHO)}
\end{align}
Here, $ L_n^{l + \frac{1}{2}} $ is the associated Laguerre polynomial and  $Y_{lm}$ is the spherical harmonic function. Refs.~ \cite{Barnes:2002mu,Ackleh:1996yt,Barnes:1996ff} explored light meson decays using SHO wave functions with $\beta$ in the range of 3.5–4.0 GeV. In this work, we adopt $\beta=4.0$ GeV.

The involved flavor wave functions $|I,I_3\rangle$ are
\begin{align}
\left| \frac{1}{2}, \frac{1}{2} \right\rangle &= \sqrt{\frac{2}{3}} \left| \Sigma_c^{(*)++} D^{(*)-} \right\rangle - \frac{1}{\sqrt{3}} \left| \Sigma_c^{(*)+} \bar{D}^{(*)0} \right\rangle, \label{flavor1}\\
\left| \frac{1}{2}, -\frac{1}{2} \right\rangle &= \frac{1}{\sqrt{3}} \left| \Sigma_c^{(*)+} D^{(*)-} \right\rangle - \sqrt{\frac{2}{3}} \left| \Sigma_c^{(*)0} \bar{D}^{(*)0} \right\rangle,\\
\left| 0, 0 \right\rangle &= \frac{1}{\sqrt{2}} \left| \Xi_c^{(\prime,*)+} D^{(*)-} \right\rangle - \frac{1}{\sqrt{2}} \left| \Xi_c^{(\prime,*)0} \bar{D}^{(*)0} \right\rangle,\\
\left| 1, 1 \right\rangle &=  \left| \Xi_c^{(\prime,*)+} \bar{D}^{(*)0} \right\rangle, \\
\left| 1, 0 \right\rangle &= \frac{1}{\sqrt{2}} \left| \Xi_c^{(\prime,*)+} D^{(*)-} \right\rangle + \frac{1}{\sqrt{2}} \left| \Xi_c^{(\prime,*)0} \bar{D}^{(*)0} \right\rangle,\\
\left| 1, -1 \right\rangle &=  \left| \Xi_c^{(\prime,*)0} D^{(*)-} \right\rangle. \label{flavor2}
\end{align}

Furthermore, the spin wave functions $|S,S_z\rangle$ of these molecular states are constructed as:
\begin{align}
    \left| S,S_z\right\rangle &=\sum_{S^{B}_z, S^{M}_z}C_{S^{B} S^{B}_z, S^{M} S^{M}_z}^{S,S_z}\left| S^{B},S^{B}_z\right\rangle\left| S^{M},S^{M}_z\right\rangle.
\end{align}
Here, the superscript ${B}$ and the superscript ${M}$denote the spins of the baryon and meson constituents, respectively. The subscript $z$ indicates the third component of the spin. The total spin and its third component for the molecular state are $S$ and $S_z$, respectively.

\section{Numerical results}\label{sec3}

\begin{figure}[!htbp]
    \centering
\includegraphics[width=1.0\linewidth]{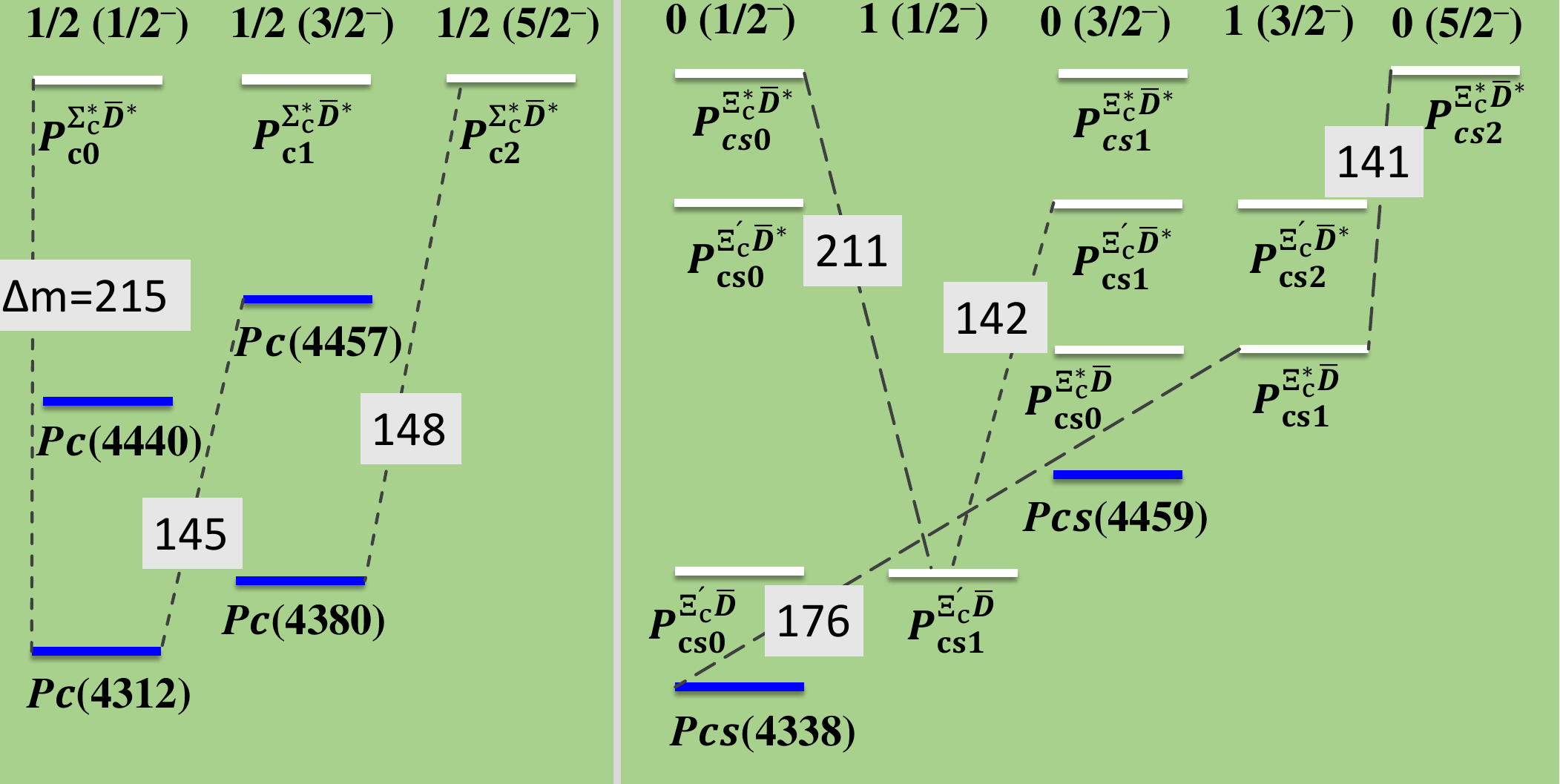}
    \caption{Mass positions for possible (strange) hidden-charm molecular states. Here, the mass gaps are in units of MeV.}
    \label{mass}
\end{figure}

Before producing our calculations, we would like to provide a brief introduction to the (strange) hidden-charm molecular pentaquarks discussed in this work. In Refs. \cite{Chen:2019asm, Chen:2022onm}, we used one-boson-exchange (OBE) effective potentials to study the mass spectrum of such states, composed of a charmed baryon and an anti-charmed meson. Our analysis successfully reproduced the masses of the $P_c(4312)$, $P_c(4440)$, $P_c(4457)$, and $P_{cs}(4459)$ within the meson-baryon molecular picture but also suggested that the $P_c(4380)$ observed in 2015 \cite{LHCb:2015yax} can be identified as a coupled-channel molecule of $\Sigma_c^*\bar{D}/\Sigma_c\bar{D}^*/\Sigma_c^*\bar{D}^*$. In this interpretation, the probability for the $\Sigma_c^*\bar{D}$ channel exceeds 87\%. Furthermore, we predicted a series of other possible (strange) hidden-charm molecular pentaquarks. The mass positions of these states are shown in Figure \ref{mass}. For convenience, we label them as follows:
\begin{eqnarray}
&&P_{c0}^{\Sigma_c^*\bar{D}^*} \sim {\Sigma_c^*\bar{D}^*}[1/2(1/2^-)],
\quad\quad
P_{cs0}^{\Xi_c^*\bar{D}^*} \sim \Xi_c^*\bar{D}^*[0(1/2^-)], \nonumber\\
&&P_{c1}^{\Sigma_c^*\bar{D}^*} \sim {\Sigma_c^*\bar{D}^*}[1/2(3/2^-)],
\quad\quad
 P_{cs1}^{\Xi_c^*\bar{D}^*} \sim \Xi_c^*\bar{D}^*[0(3/2^-)], \nonumber\\
&&P_{c2}^{\Sigma_c^*\bar{D}^*} \sim {\Sigma_c^*\bar{D}^*}[1/2(5/2^-)],
\quad\quad P_{cs2}^{\Xi_c^*\bar{D}^*} \sim \Xi_c^*\bar{D}^*[0(5/2^-)],\nonumber\\
&&P_c(4312) \sim \Sigma_c\bar{D}/\Sigma_c\bar{D}^*/\Sigma_c^*\bar{D}^*[1/2(1/2^-)],\nonumber\\
&&P_c(4380) \sim \Sigma_c^*\bar{D}/\Sigma_c\bar{D}^*/\Sigma_c^*\bar{D}^*[1/2(3/2^-)],\nonumber\\
&&P_c(4440) \sim \Sigma_c\bar{D}^*/\Sigma_c^*\bar{D}^*[1/2(1/2^-)],\nonumber\\
&&P_c(4457) \sim \Sigma_c\bar{D}^*/\Sigma_c^*\bar{D}^*[1/2(3/2^-)],\nonumber\\
&&P_{cs}(4338) \sim \Xi_c\bar{D}/\Xi_c^{\prime}\bar{D}/\Xi_c\bar{D}^*/\Xi_c^*\bar{D}/\Xi_c^{\prime}\bar{D}^*/\Xi_c^*\bar{D}^*[0(1/2^-)], \nonumber\\
&&P_{cs}(4459) \sim \Xi_c\bar{D}^*/\Xi_c^*\bar{D}/\Xi_c^{\prime}\bar{D}^*/\Xi_c^*\bar{D}^*[0(3/2^-)], \nonumber\\
&&P_{cs0}^{\Xi_c^{\prime}\bar{D}} \sim \Xi_c^{\prime}\bar{D}/\Xi_c\bar{D}^*/\Xi_c^*\bar{D}/\Xi_c^{\prime}\bar{D}^*/\Xi_c^*\bar{D}^*[0(1/2^-)], \nonumber\\
&&P_{cs1}^{\Xi_c^{\prime}\bar{D}} \sim \Xi_c^{\prime}\bar{D}/\Xi_c\bar{D}^*/\Xi_c^*\bar{D}/\Xi_c^{\prime}\bar{D}^*/\Xi_c^*\bar{D}^* [1(1/2^-)], \nonumber\\
&&P_{cs0}^{\Xi_c^*\bar{D}} \sim \Xi_c^*\bar{D}/\Xi_c^{\prime}\bar{D}^*/\Xi_c^*\bar{D}^*[0(3/2^-)], \nonumber\\
&&P_{cs1}^{\Xi_c^*\bar{D}} \sim \Xi_c^*\bar{D}/\Xi_c^{\prime}\bar{D}^*/\Xi_c^*\bar{D}^*[1(3/2^-)], \nonumber\\
&&P_{cs0}^{\Xi_c^{\prime}\bar{D}^*} \sim \Xi_c^{\prime}\bar{D}^*/\Xi_c^*\bar{D}^*[0(1/2^-)], \nonumber\\
&&P_{cs1}^{\Xi_c^{\prime}\bar{D}^*} \sim \Xi_c^{\prime}\bar{D}^*/\Xi_c^*\bar{D}^*[0(3/2^-)], \nonumber\\
&&P_{cs2}^{\Xi_c^{\prime}\bar{D}^*} \sim \Xi_c^{\prime}\bar{D}^*/\Xi_c^*\bar{D}^*[1(3/2^-)]. \nonumber
\end{eqnarray}

Our findings also highlight the important role of coupled-channel effects in forming the $P_c(4312)$,  $P_c(4457)$, and $P_{cs}(4459)$ states. Specifically: 
\begin{itemize}
    \item For $P_c(4312)$, the probabilities for the $\Sigma_c\bar{D}$, $\Sigma_c\bar{D}^*$, and $\Sigma_c^*\bar{D}^*$ channels are 0.66, 0.18, and 0.16, respectively.
    \item For $P_c(4457)$, the probabilities for the $\Sigma_c\bar{D}^*$ and $\Sigma_c^*\bar{D}^*$ channels are 0.75 and 0.25, respectively.
    \item For $P_{cs}(4459)$, the probabilities for the $\Xi_c\bar{D}^*$, $\Xi_c^*\bar{D}$, $\Xi_c^{\prime}\bar{D}$, $\Xi_c^*\bar{D}^*$ channels are 0.39, 0.35, 0.07, and 0.19, respectively.
\end{itemize}

In the following section, we present the results for the partial widths of pion emission decays between these possible (strange) hidden-charm molecular pentaquarks. ased on the conservation of quantum numbers and the constraints of phase space, we will discuss the following decay processes for hidden-charm molecules:
\begin{itemize}
    \item $P_c(4457) \to P_c(4312)+\pi$,
    \item $P_c^{\Sigma_c^*\bar{D}^*} \to P_c(4312)+\pi$,
    \item $P_c^{\Sigma_c^*\bar{D}^*} \to P_c(4380)+\pi$.
\end{itemize}

\subsection{$P_c(4457) \to P_c(4312)+\pi$}

For the $P_c(4457) \to P_c(4312)+\pi$ process, we input the wave functions from our previous work \cite{Chen:2019asm}. The calculated decay widths for the charged and neutral pion emission are 7.92 keV and 9.70 keV, respectively. According to the flavor wave functions in Eqs. (\ref{flavor1})-(\ref{flavor2}), the decay amplitudes obey the relation $\mathcal{M}(P_c(4457) \to P_c(4312)+\pi^{\pm}) = \sqrt{2}\mathcal{M}(P_c(4457) \to P_c(4312)+\pi^{0})$. The similar magnitudes of the decay widths, despite this amplitude relation, result from the slightly different phase spaces for the charged and neutral pion emission. When coupled-channel effects are included, pion radiation from both the charmed baryon and the anti-charmed meson constituents contribute to the total decay width:
\begin{eqnarray}
\mathcal{M}_{\text{Total}} &=& \mathcal{M}(\Sigma_c^{(*)}\to\Sigma_c^{(*)}+\pi) + \mathcal{M}(\bar{D}^{(*)}\to\bar{D}^{(*)}+\pi).\quad
\end{eqnarray} 

Our results show that the partial decay widths for $\pi^0$ emission originating solely from the charmed baryon and anti-charmed meson transitions are 5.56 keV and 0.57 keV, respectively. This indicates that the pion emission in the decay $P_c(4457)\to P_c(4312)$ is dominated by the charmed baryon components. Furthermore, the total amplitude arises from a coherent sum of the contributions from both constituent hadrons. Since pion emission from the charmed baryon is absent in a pure single-channel picture, we conclude that coupled-channel effects play a crucial role in this decay.

To further investigate the role of coupled-channel effects, we also calculated the pion emission widths for $P_c(4457) \to P_c(4312)+\pi$ within a single-channel scenario, where the decay proceeds only via the $\bar{D}^*\to\bar{D}+\pi$ interaction. In this case, $P_c(4457)$ and $P_c(4312)$ are assigned as a pure $\Sigma_c\bar{D}^*$ molecule with $1/2(3/2^-)$ and a pure $\Sigma_c\bar{D}$ molecule with $1/2(1/2^-)$, respectively. In Ref. \cite{Ling:2021lmq}, authors investigated the processes of the $P_c(4457) \to P_c(4312)+\pi$ through the triangle mechanism  within a single-channel scenario. We approximate the wave functions of these loosely bound states using $S-$wave Gauss functions \cite{Wang:2019spc} of the form $\phi(p)=\text{exp}(-\bm{p}^2/2\beta_1^2)/(\beta_1\pi)^{3/2}$, where $\bm{p}=(m_M{\bm{p}}_M-m_B{\bm{p}}_B)/(m_M+m_B)$ is the relative momentum. Here, $m_{M(B)}$ and $\bm{p}_{M(B)}$ are the masses and momenta of the meson (baryon) constituents, and $\beta_1=\sqrt{3\mu(m_M+m_B-m_{P_c})}$ is the oscillator parameter with the reduced mass $\mu=m_Mm_B/(m_M+m_B)$. The resulting decay widths for $P_c(4457) \to P_c(4312)+\pi^0$ and $P_c(4457) \to P_c(4312)+\pi^+$ are 3.39 keV and 2.79 keV, respectively. These values are somewhat smaller than the decay widths of a free $D^*$ meson  \cite{CLEO:1997rew,BaBar:2013thi}, primarily due to amplitude cancellation between the $\Sigma_cD^{*-}\to \Sigma_cD^-+\pi^0$ and $\Sigma_c\bar{D}^{(*0)}\to \Sigma_c\bar{D}^0+\pi^0$ transitions. 

We also calculated the decay widths for $P_c(4457) \to P_c(4312)+\pi$ with $P_c(4457)$ assigned as a pure $\Sigma_c\bar{D}^*$ molecule with $1/2(1/2^-)$ and $P_c(4312)$ as a $\Sigma_c\bar{D}$ molecule with $1/2(1/2^-)$. In this configuration, the obtained decay widths for charged and neutral pion emission are 0.10 keV and 0.11 keV, respectively, which are significantly smaller than those for the $J^P=3/2^-$ assignment of the $P_c(4457)$.

In summary, we find the following order of magnitude for the pion emission widths in the $P_c(4457) \to P_c(4312)+\pi$ process:
\begin{eqnarray}
\Gamma(1/2^-):\Gamma(3/2^-):\Gamma^{\text{C}}(3/2^-)=0.10:3.00:9.00, 
\end{eqnarray}
where $\Gamma(1/2^-)$ and $\Gamma(3/2^-)$ denote the decay widths with  $P_c(4457)$ as the pure $\Sigma_c\bar{D}^*$ molecule with $I(J^P)=1/2(1/2^-)$ and $1/2(3/2^-)$, respectively, and $\Gamma^{\text{C}}(3/2^-)$ represents the decay width with both $P_c(4457)$ and $P_c(4312)$ treated as coupled-channel molecules \cite{Chen:2019asm}. These results provide important and helpful insights for understanding the internal structures and quantum number assignments of the $P_c(4457)$ and $P_c(4312)$ states.

\subsection{$P_c^{\Sigma_c^*\bar{D}^*}\to P_c(4312)+\pi$}

We next compute the decay widths for $\Sigma_c^*\bar{D}^*$ molecules with $I(J^P)=1/2(1/2^-, 3/2^-, 5/2^-)$ decaying into $P_c(4312)+\pi$. For initial states with $I(J^P)=1/2(1/2^-, 3/2^-)$, the decay proceeds via $P-$wave interactions. In a single-channel analysis where $P_c(4312)$ is treated as a pure
$\Sigma_c\bar{D}$ molecule with $1/2(1/2^-)$, these pion emission processes are forbidden due to the absence of common configurations between the initial and final molecular states.

\begin{figure}[!htbp]
    \centering
\includegraphics[width=1\linewidth]{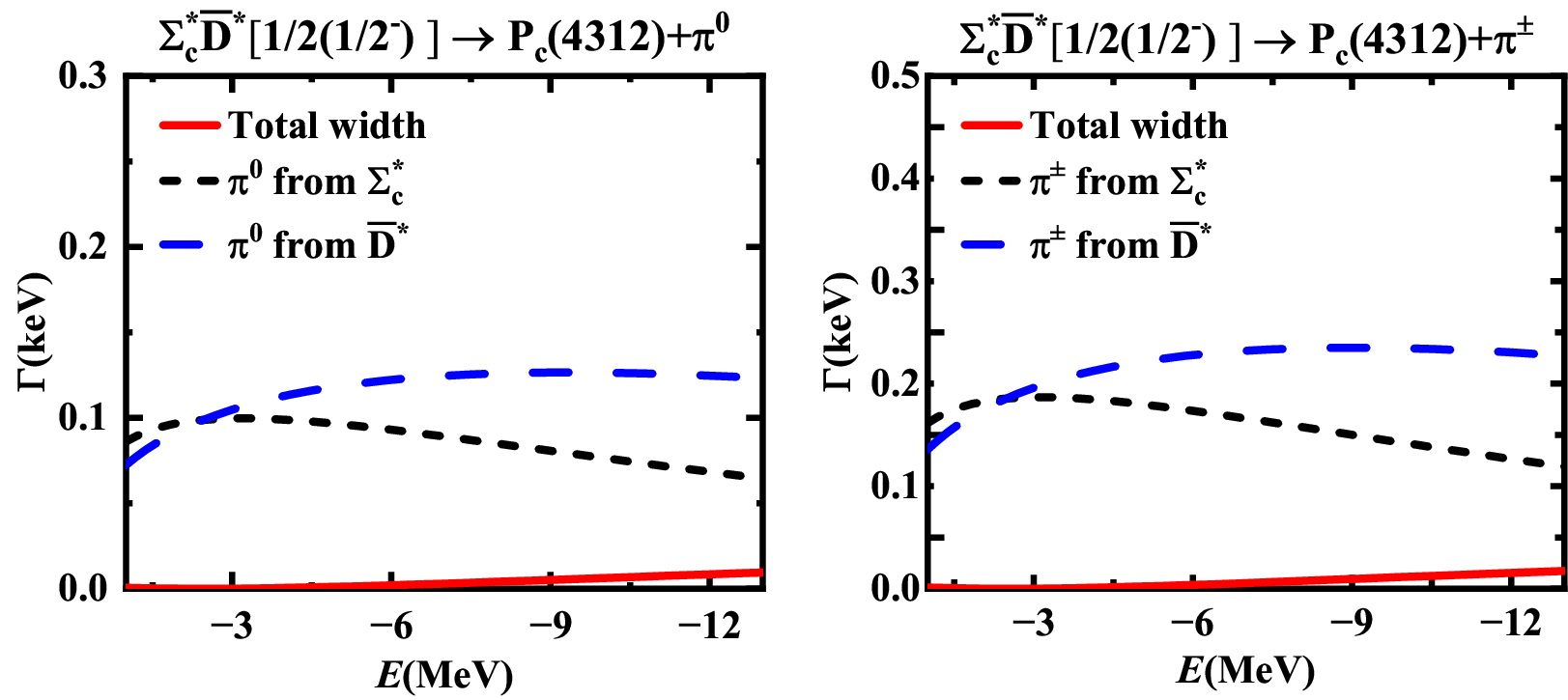}
\includegraphics[width=1\linewidth]{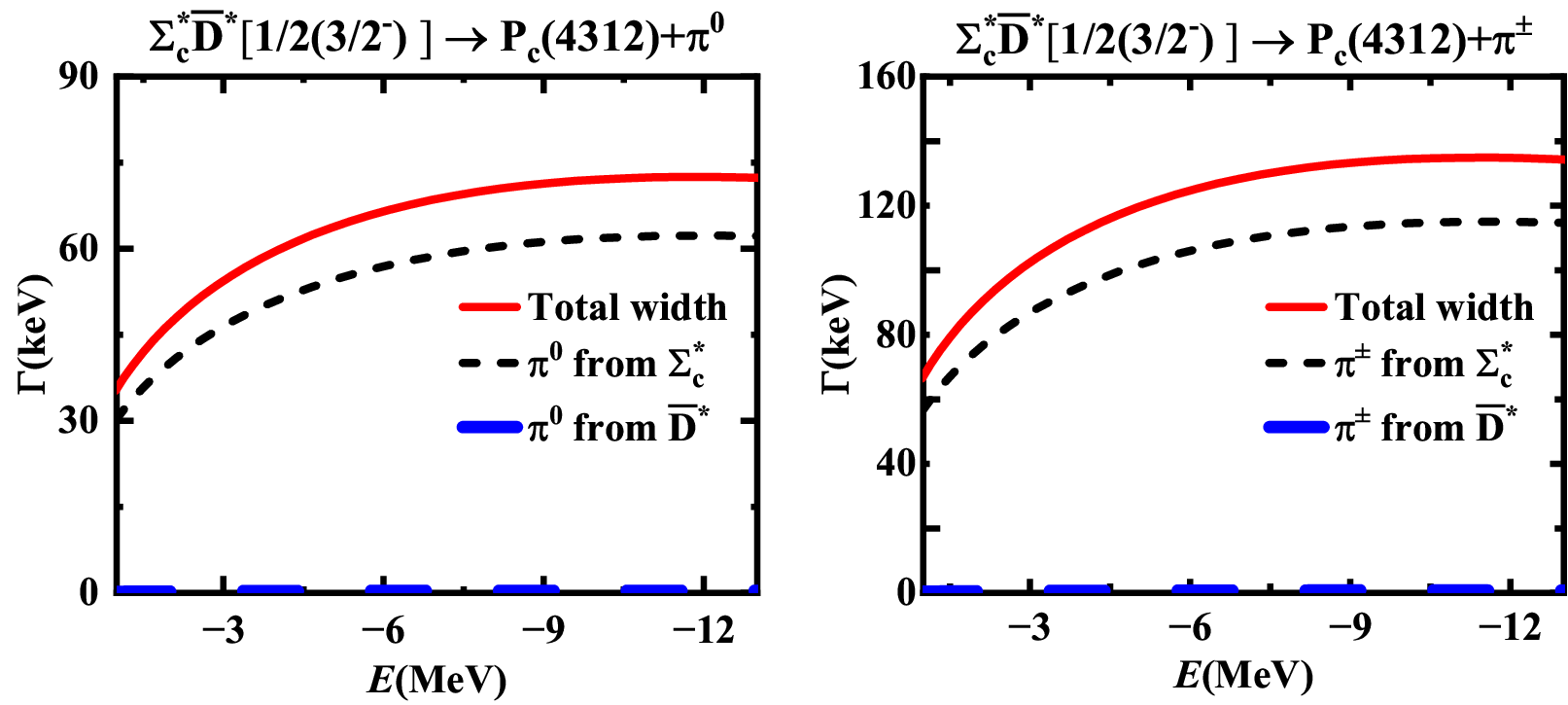}
    \caption{The pion emission widths for $\Sigma_c^*\bar{D}^*$ molecules with $I(J^P)=1/2(1/2^-, 3/2^-)$ decaying into $P_c(4312)$.}
    \label{fig1}
\end{figure}

When coupled-channel effects are included, as illustrated in Figure \ref{fig1}, the situation changes. Our results show that for the decay $P_c^{\Sigma_c^*\bar{D}^*}[1/2(1/2^-)]\to P_c(4312)+\pi^{\pm,0}$, the partial widths are less than 0.1 keV for binding energies $E$ ranging from 0 to $-$12 MeV. These small widths are dominated by pion emission from the anti-charmed meson components, $\bar{D}^{(*)}\to\bar{D}^{(*)}+\pi$. The contribution from pion emission off the charmed baryon components is negligible. In line with the isospin factor relations, the decay width for neutral pion emission is approximately a factor of two smaller than for charged pion emission.

For the decay $P_c^{\Sigma_c^*\bar{D}^*}[1/2(3/2^-)]\to P_c(4312)+\pi^{\pm}$, the partial width ranges from 60 to 140 keV for initial state binding energies $E>-15$ MeV. In this case, pion emission from the charmed baryon components plays a dominant role. Consistent with the isospin relation $\mathcal{M}(P_c^i \to P_c^f+\pi^{\pm}) = \sqrt{2}\mathcal{M}(P_c^i \to P_c^f+\pi^{0})$, the neutral pion emission width is again a factor of two smaller than its charged counterpart.

Finally, for the decay $P_c^{\Sigma_c^*\bar{D}^*}[1/2(5/2^-)]\to P_c(4312)+\pi$, the process must proceed via an $F-$wave interaction. This higher partial wave is expected to be strongly suppressed compared to the $P-$wave decays discussed above.

\subsection{ $P_c^{\Sigma_c^*\bar{D}^*} \to P_c(4380)+\pi$}

\begin{figure}[!htbp]
    \centering
\includegraphics[width=1\linewidth]{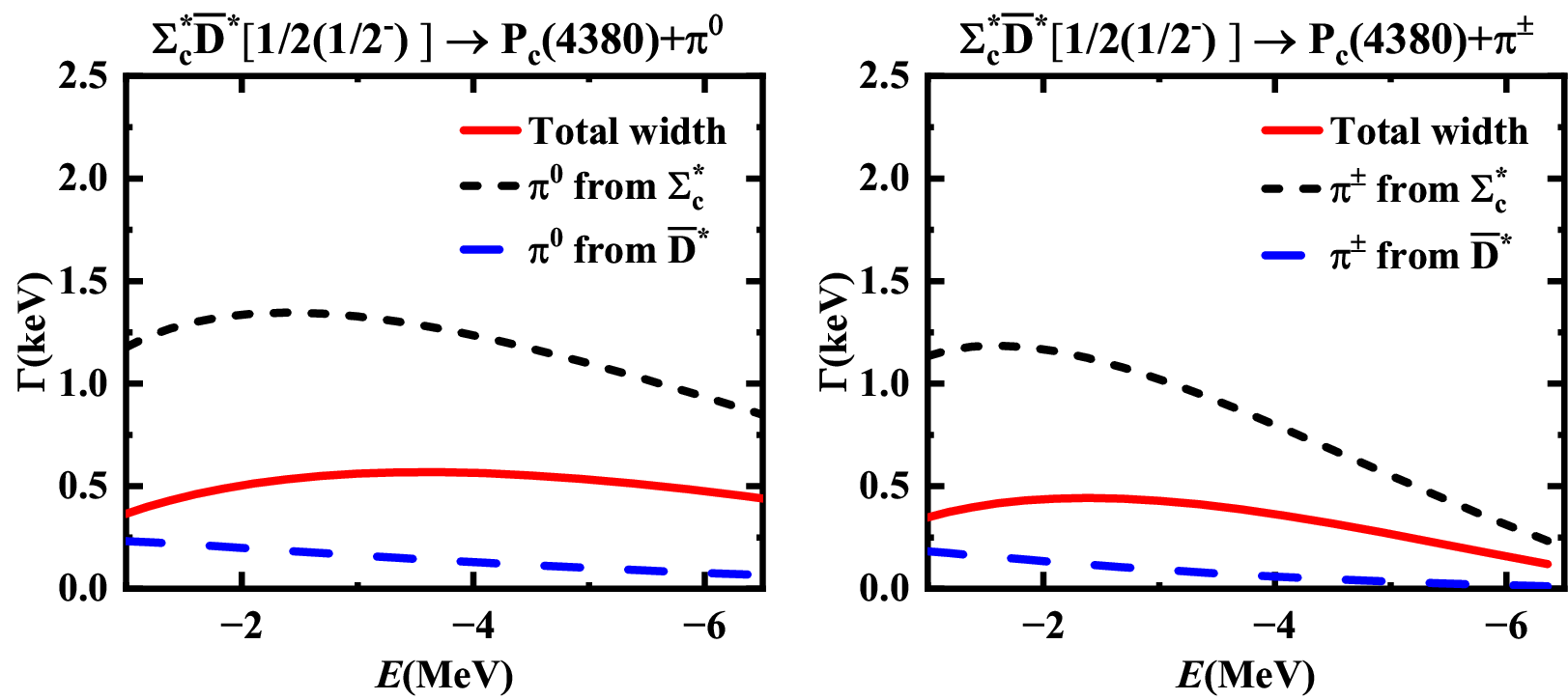}
\includegraphics[width=1\linewidth]{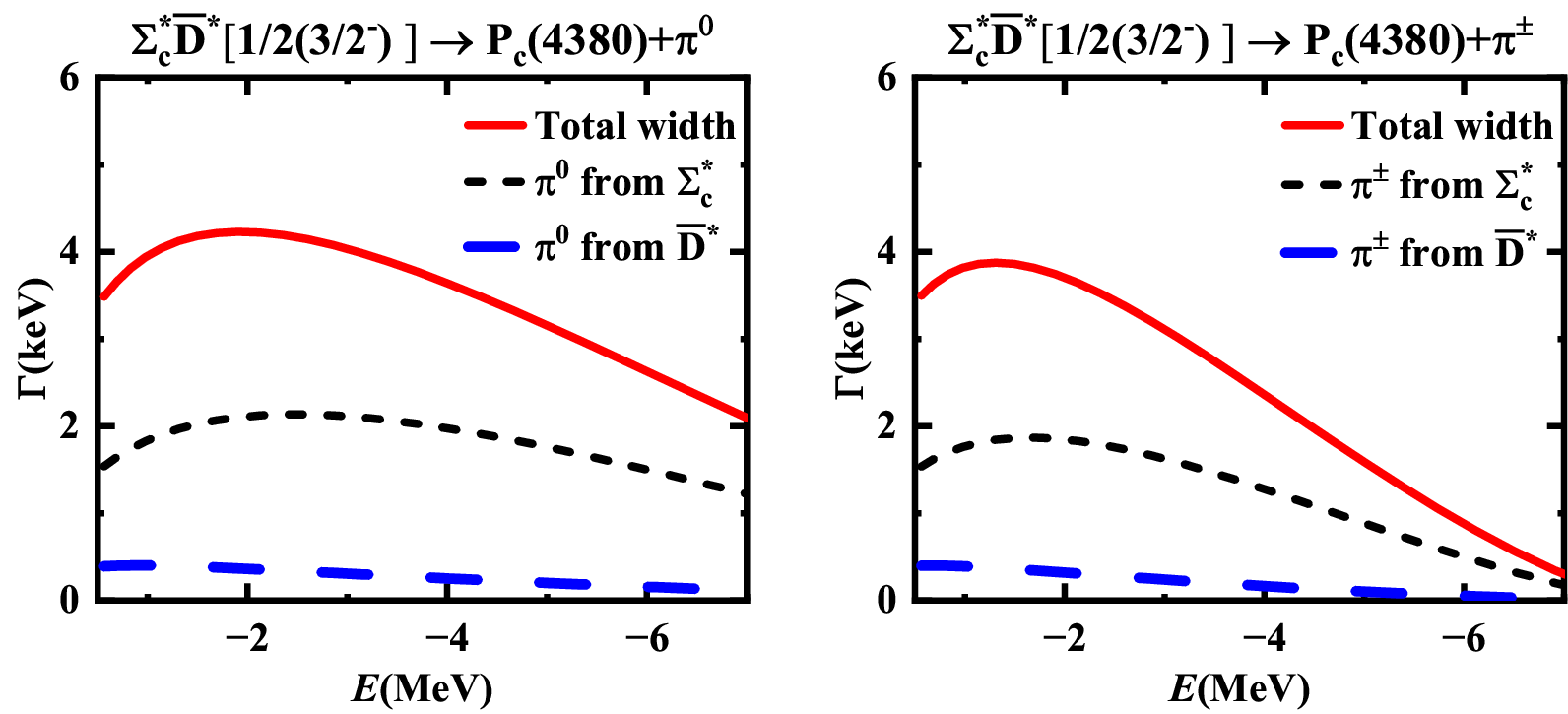}
\includegraphics[width=1\linewidth]{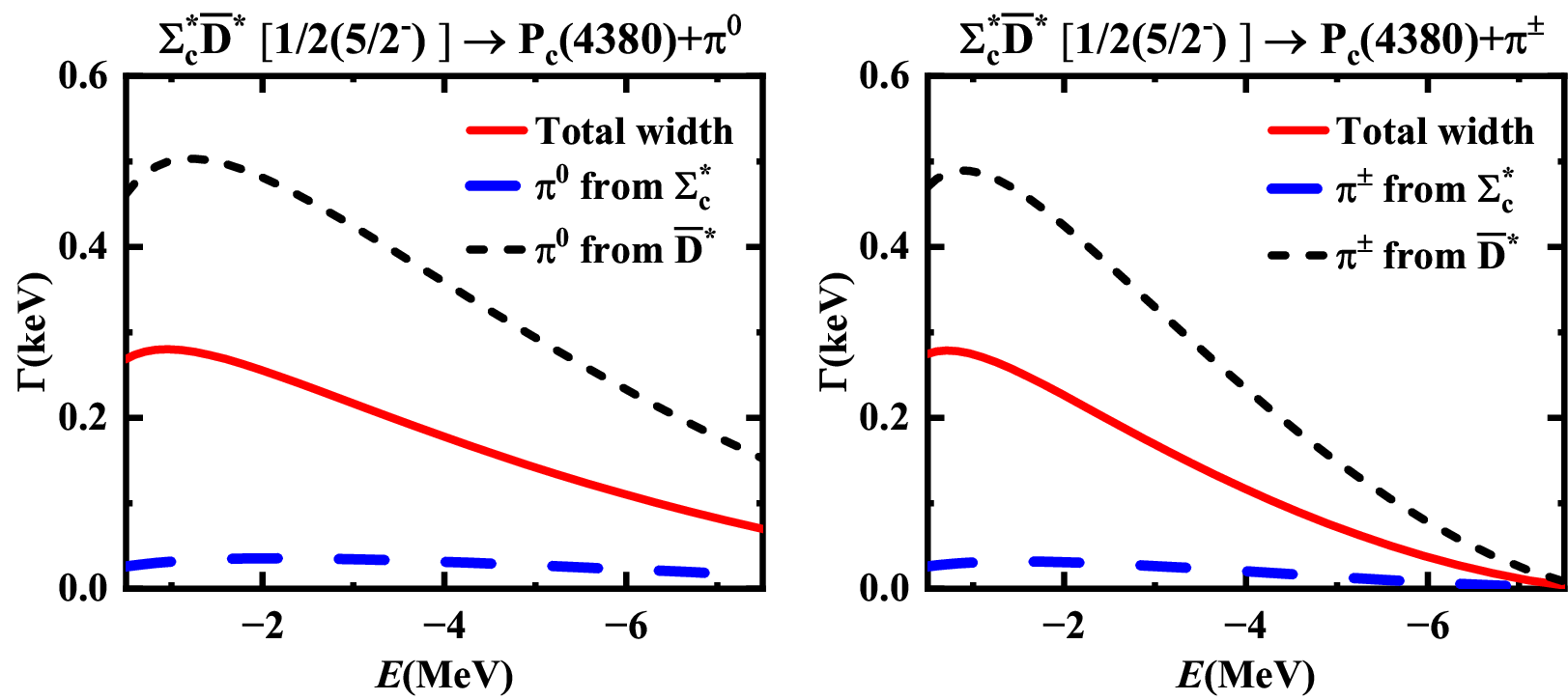}
        \caption{The pion emission widths for the $\Sigma_c^*\bar{D}^*$ molecules with $I(J^P)=1/2(1/2^-, 3/2^-, 5/2^-)$ decaying into $P_c(4380)$.}
    \label{to4380}
\end{figure}

For the decay processes $P_c^{\Sigma_c^*\bar{D}^*} \to P_c(4380)+\pi$, the available phase space is relatively limited compared to the decays to $P_c(4312)+\pi$. This significantly influences the resulting decay widths, as shown in Figure \ref{to4380}. The limited phase space manifests in two key aspects: first, the decay widths for neutral and charged pion emission become comparable, despite the isospin relation $\mathcal{M}(P_c^i \to P_c^f+\pi^{\pm}) = \sqrt{2}\mathcal{M}(P_c^i \to P_c^f+\pi^{0})$; second, the decay widths decrease with increasing binding energy of the $\Sigma_c^*\bar{D}^*$ molecules.

We first discuss the decay properties assuming $P_c(4380)$ as a pure $\Sigma_c^*\bar{D}$ molecule with $1/2(3/2^-)$. In this scenario, pion emission occurs via the transition between the anti-charmed meson components, $\bar{D}^*\to \bar{D}+\pi$. Since the charmed baryon components act as spectators, the decay width is independent of the total spin of the initial $\Sigma_c^*\bar{D}^*$ state. Our calculations show decay widths on the order of several keV for initial states with $J^P=1/2^-, 3/2^-$, and $5/2^-$.

In the coupled channel case, where $P_c(4380)$ is a mixture of $\Sigma_c^*\bar{D}$, $\Sigma_c\bar{D}^*$, and $\Sigma_c^*\bar{D}^*$ with $1/2(3/2^-)$, pion emission from the charmed baryon components also contributes to the total decay width. Consequently, the decay widths for initial $\Sigma_c^*\bar{D}^*$ states with different spin-parities differ substantially. Figure \ref{to4380} shows the dependence of these decay widths on the binding energy.

For the process $P_c^{\Sigma_c^*\bar{D}^*}[1/2(1/2^-)]\to P_c(4380)+\pi$, the pion emission amplitude from the charmed baryon component is stronger than that from the anti-charmed meson component. However, these two amplitudes interfere destructively, resulting in a very small total decay width of less than 0.5 keV.

In contrast, for the process $P_c^{\Sigma_c^*\bar{D}^*}[1/2(3/2^-)]\to P_c(4380)+\pi$, the individual amplitudes from the meson and baryon components are of similar magnitude but interfere constructively.This leads to a significantly larger decay width, as shown in Figure \ref{to4380}, where both charged and neutral pion emission widths can reach a few keV for binding energies $E>-10$ MeV.

For the process $P_c^{\Sigma_c^*\bar{D}^*}[1/2(5/2^-)]\to P_c(4380)+\pi$, the total decay width is less than 1 keV. This is dominated by the $\bar{D}^*\to\bar{D}^{(*)}+\pi$ interactions, the pion emission from the charmed baryon component is very small. Furthermore, these two contributions interfere destructively.

\subsection{$P_{cs}^{i}\to P_{cs}^f+\pi$}

We next compute the pion emission widths for transitions between possible strange hidden-charm molecular pentaquarks, $P_{cs}^{i}\to P_{cs}^f+\pi$, within the coupled-channel framework. The processes discussed, constrained by quantum number conservation, are:
\begin{itemize}
    \item $P_{cs}^{\Xi_c^{*}\bar{D}}[1(3/2^-)]\to P_{cs}(4338)+\pi$,
    \item $P_{cs}^{\Xi_c^{\prime}\bar{D}^*}[1(3/2^-)] \to P_{cs}^{\Xi_c^{\prime}\bar{D}}[0,1(1/2^-)]+\pi$,
    \item $P_{cs}^{\Xi_c^{\prime}\bar{D}^*}[0(1/2^-, 3/2^-)] \to P_{cs}^{\Xi_c^{\prime}\bar{D}}[1(1/2^-)]+\pi$,
    \item $P_{cs}^{\Xi_c^*\bar{D}^*}[0(1/2^-,3/2^-)] \to P_{cs}^{\Xi_c^{\prime}\bar{D}}[1(1/2^-)]+\pi$,
    \item $P_{cs}^{\Xi_c^*\bar{D}^*}[0(1/2^-,3/2^-,5/2^-)] \to P_{cs}^{\Xi_c^*\bar{D}}[1(3/2^-)]+\pi$.
\end{itemize}

\renewcommand\tabcolsep{0.2cm}
    \renewcommand{\arraystretch}{1.7}
\begin{table}[!htbp]
    \centering
    \caption{The $\pi$ emission widths between possible strange hidden-charm molecular pentaquarks. Here, the units of the widths are keV.}
    \label{pcs}
    \begin{tabular}{l|c|c}
    \toprule[1pt]
    \toprule[1pt]
        Processes   &$\pi^0$  &$\pi^{\pm}$  \\\hline
$P_{cs1}^{\Xi_c^{*}\bar{D}}[1(3/2^-)]\to P_{cs}(4338)+\pi$    
    &$130\sim280$  &$100\sim250$ \\ 
$P_{cs1}^{\Xi_c^{\prime}\bar{D}^*}[1(3/2^-)] \to P_{cs0}^{\Xi_c^{\prime}\bar{D}}[0(1/2^-)]+\pi$ 
    &4.0   &0.6    \\
$P_{cs1}^{\Xi_c^{\prime}\bar{D}^*}[1(3/2^-)] \to P_{cs1}^{\Xi_c^{\prime}\bar{D}}[1(1/2^-)]+\pi$   
    &$25\sim50$  &$ 3\sim8$  \\
$P_{cs0}^{\Xi_c^{\prime}\bar{D}^*}[0(1/2^-)] \to P_{cs1}^{\Xi_c^{\prime}\bar{D}}[1(1/2^-)]+\pi$ 
    &$1.5$  &$ 0.2$ \\
$P_{cs1}^{\Xi_c^{\prime}\bar{D}^*}[0(3/2^-)] \to P_{cs1}^{\Xi_c^{\prime}\bar{D}}[1(1/2^-)]+\pi$ &$0.8$  &$ 0.1$  \\
$P_{cs0}^{\Xi_c^*\bar{D}^*}[0(1/2^-)] \to P_{cs1}^{\Xi_c^{\prime}\bar{D}}[1(1/2^-)]+\pi$ &$30\sim100$  &$ 20\sim90$\\
$P_{cs1}^{\Xi_c^*\bar{D}^*}[0(3/2^-)] \to P_{cs1}^{\Xi_c^{\prime}\bar{D}}[1(1/2^-)]+\pi$ &$50\sim150$  &$ 30\sim150$\\
$P_{cs0}^{\Xi_c^*\bar{D}^*}[0(1/2^-)] \to P_{cs1}^{\Xi_c^*\bar{D}}[1(3/2^-)]+\pi$ &$0.5$  &$0.1$ \\
$P_{cs1}^{\Xi_c^*\bar{D}^*}[0(3/2^-)] \to P_{cs1}^{\Xi_c^*\bar{D}}[1(3/2^-)]+\pi$ &$1.7$  &$0.3$ \\
$P_{cs2}^{\Xi_c^*\bar{D}^*}[0(5/2^-)] \to P_{cs1}^{\Xi_c^*\bar{D}}[1(3/2^-)]+\pi$ &$2.8$  &$0.4$  \\\bottomrule[1pt]\bottomrule[1pt]
    \end{tabular}
\end{table}

The corresponding decay widths are summarized in Table \ref{pcs}. In these calculations, the binding energies for both the initial and final molecular states were varied within the range $E>-10$ MeV, while the binding energy of $P_{cs}(4338)$ was fixed at $E=-1.06$ MeV. Our findings are as follows:
\begin{enumerate}
    \item For the process $P_{cs1}^{\Xi_c^{*}\bar{D}}[1(3/2^-)]\to P_{cs}(4338)+\pi$, the partial widths for neutral and charged pion emission are very similar. This occurs despite the isospin relation for the decay amplitude, $\mathcal{M}(P_{cs}^i \to P_{cs}^f+\pi^{\pm}) = -\mathcal{M}(P_{cs}^i \to P_{cs}^f+\pi^{0})$, due to nearly identical phase space for both final states. When the initial molecular state has a binding energy $E>-10$ MeV, the decay widths range from 130 to 280 keV for $\pi^0$ and from 100 to 250 keV for $\pi^{\pm}$. These decays proceed exclusively via pion emission from the charmed baryon components, $\Xi_c^{(',*)}\to\Xi_c^{(',*)}+\pi$. 
    \item For the process $P_{cs1}^{\Xi_c^{\prime}\bar{D}^*}[1(3/2^-)] \to P_{cs0}^{\Xi_c^{\prime}\bar{D}}[0(1/2^-)]+\pi$, the partial widths are only a few keV, being strongly suppressed by the limited phase space. The dominant contributions are attributed to interactions involving the charmed baryons, $\Xi_c^{(',*)}\to\Xi_c^{(',*)}+\pi$.
    \item For the process $P_{cs1}^{\Xi_c^{\prime}\bar{D}^*}[1(3/2^-)] \to P_{cs1}^{\Xi_c^{\prime}\bar{D}}[1(1/2^-)]+\pi$, the partial widths can reach several tens of keV, significantly larger than the previous case. This enhancement results from constructive interference among the amplitudes from different coupled channels. Pion emission from the charmed baryon components remains the dominant mechanism.
    \item For the following processes, the partial widths are very small, typically only a few keV:
    \begin{itemize}
        \item $P_{cs0}^{\Xi_c^{\prime}\bar{D}^*}[0(1/2^-)] \to P_{cs1}^{\Xi_c^{\prime}\bar{D}}[1(1/2^-)]+\pi$,
        \item $P_{cs1}^{\Xi_c^{\prime}\bar{D}^*}[0(3/2^-)] \to P_{cs1}^{\Xi_c^{\prime}\bar{D}}[1(1/2^-)]+\pi$,
        \item $P_{cs0}^{\Xi_c^*\bar{D}^*}[0(1/2^-)] \to P_{cs1}^{\Xi_c^{\prime}\bar{D}}[1(1/2^-)]+\pi$,
        \item $P_{cs1}^{\Xi_c^*\bar{D}^*}[0(3/2^-)] \to P_{cs1}^{\Xi_c^*\bar{D}}[1(3/2^-)]+\pi$,
        \item $P_{cs2}^{\Xi_c^*\bar{D}^*}[0(5/2^-)] \to P_{cs1}^{\Xi_c^*\bar{D}}[1(3/2^-)]+\pi$.
    \end{itemize}
    The decay widths are primarily due to pion emission between the anti-charmed meson components, $\bar{D}^{(*)}\to\bar{D}^{(*)}+\pi$. The extremely limited phase space leads to significant differences between the neutral and charged pion emission widths, even though the isospin relation $\mathcal{M}(P_{cs}^i \to P_{cs}^f+\pi^{\pm}) = -\mathcal{M}(P_{cs}^i \to P_{cs}^f+\pi^{0})$ holds. 
    \item For the processes $P_{cs0}^{\Xi_c^*\bar{D}^*}[0(1/2^-)] \to P_{cs1}^{\Xi_c^{\prime}\bar{D}}[1(1/2^-)]+\pi$ and $P_{cs1}^{\Xi_c^*\bar{D}^*}[0(3/2^-)] \to P_{cs1}^{\Xi_c^{\prime}\bar{D}}[1(1/2^-)]+\pi$ processes, pion emission from the charmed baryon components plays a significant role. The partial widths are around several tens to one hundred keV for $E>-10$ MeV, and the widths for neutral and charged pion emission are again comparable.
\end{enumerate}

\section{Summary}\label{sec4}

\begin{figure*}[!htbp]
    \centering
\includegraphics[width=0.75\linewidth]{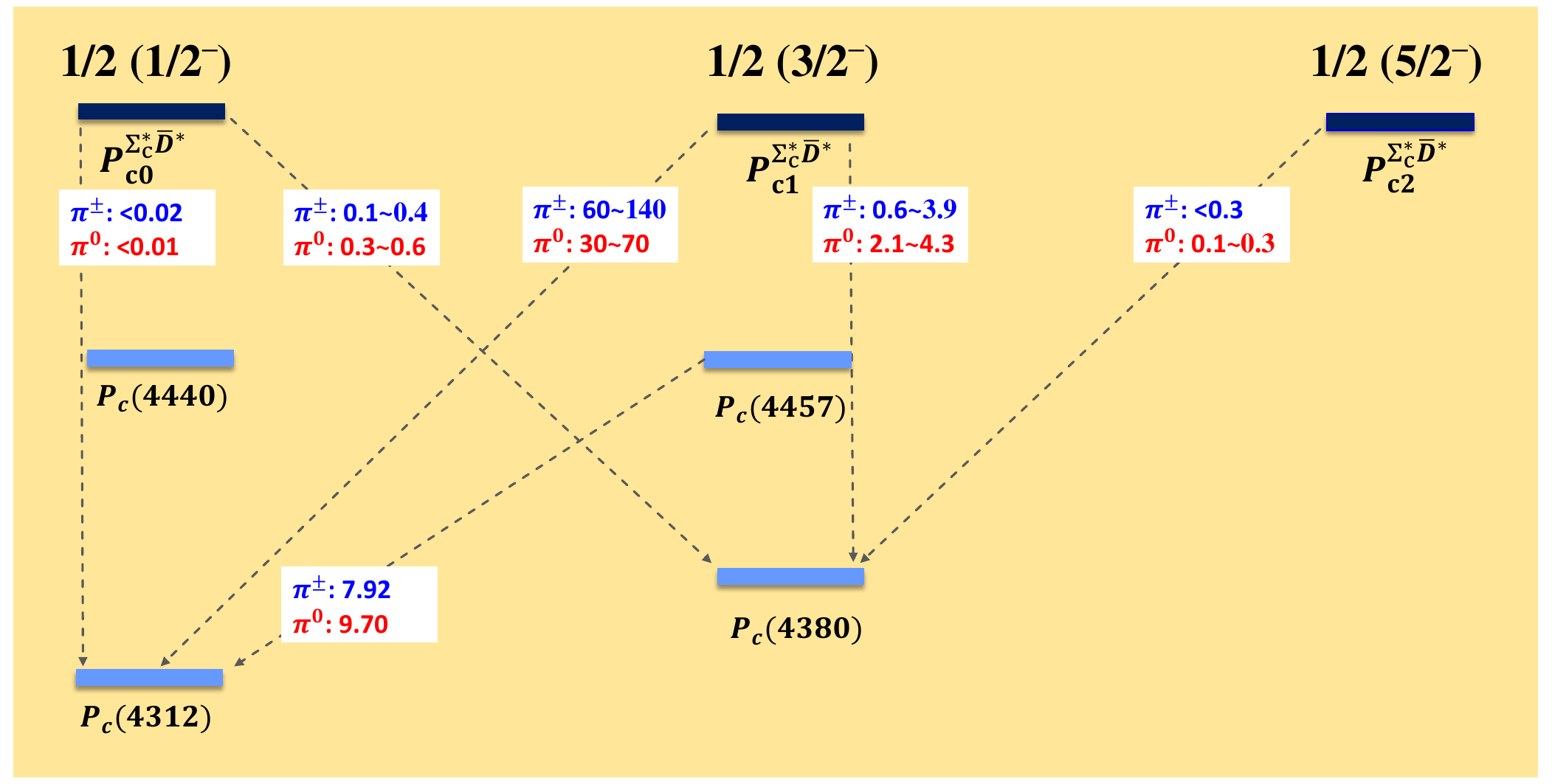}
\includegraphics[width=0.75\linewidth]{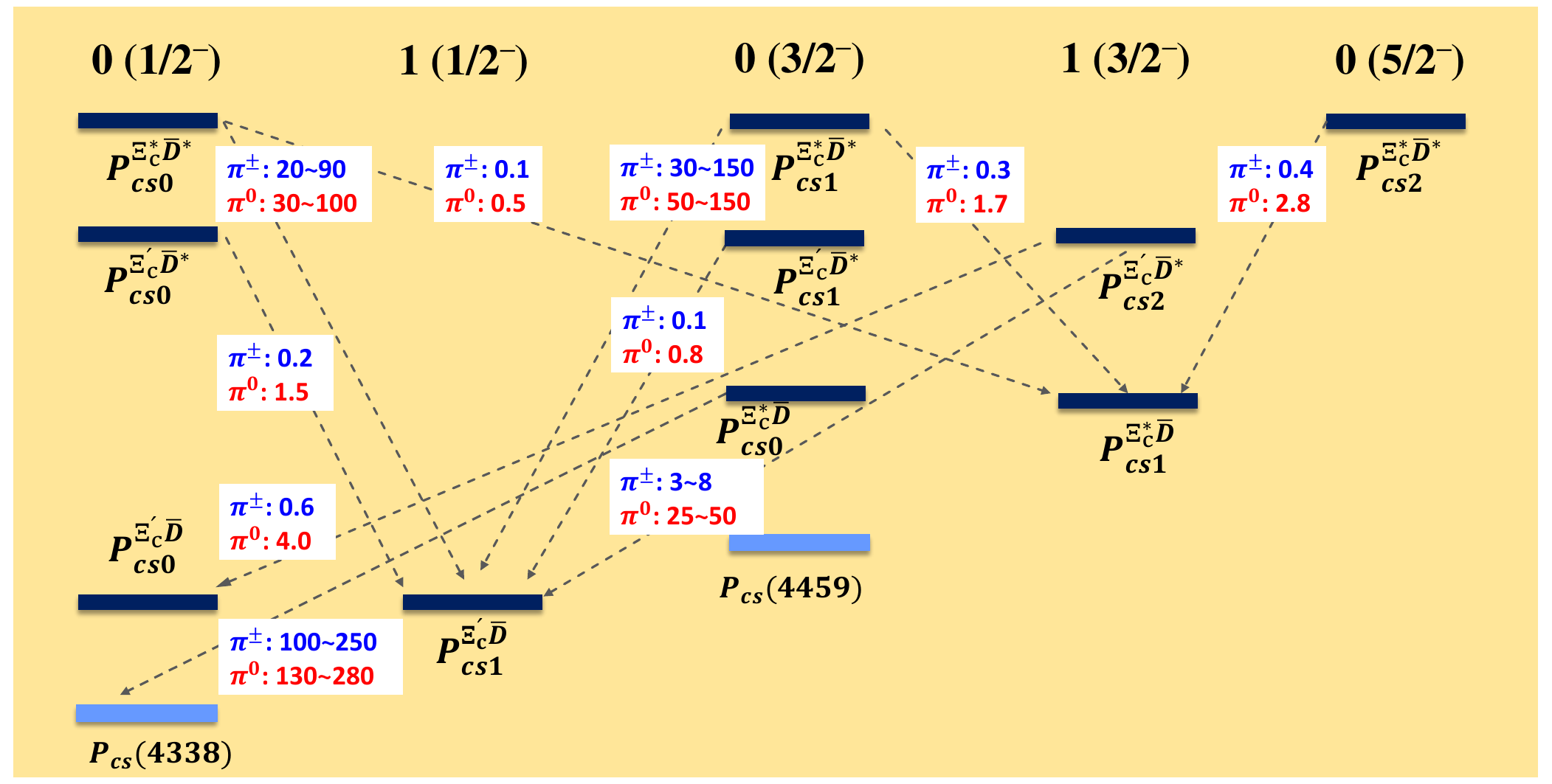}
    \caption{A summary of the pion emission properties of possible $P_c/P_{cs}$ molecules. Here, the decay widths are in the unit of keV.}
    \label{figsummary}
\end{figure*}

The discovery of hidden-charm pentaquarks constitutes a milestone in hadron physics, not only by confirming a new form of matter but, more importantly, by opening a new frontier for probing the strong interaction in the non-perturbative regime. Despite extensive theoretical and experimental progress over the past decade, fundamental questions regarding their internal structure and quantum numbers remain actively debated.

In this context, the study of light meson emission decays emerges as a powerful and independent diagnostic tool. These radiative processes are highly sensitive to the strong interactions at play and are intimately linked to the spatial wave functions and quantum numbers of the involved hadrons, thereby providing crucial information to discriminate between different structural models.

In this work, we have investigated the pion emission properties of $P_c/P_{cs}$ states within a molecular scenario, focusing on both experimentally observed states \cite{LHCb:2015yax, LHCb:2019kea, LHCb:2020jpq, LHCb:2022ogu} and our own theoretical predictions \cite{Chen:2019asm, Chen:2022onm}. Our framework is based on the chiral quark model, incorporating coupled-channel effects to derive the pion emission interactions. The numerical calculations utilize the spatial wave functions obtained from our previous dynamical studies \cite{Chen:2019asm, Chen:2022onm}.

In Figure \ref{figsummary}, we present the pion emission widths between possible $P_c/P_{cs}$ molecules. Our key findings are as follows:
\begin{itemize}
    \item The calculated pion emission widths exhibit a strong dependence on the quantum number configurations and the available phase space.
    \item Coupled-channel effects play an essential role in determining the decay amplitudes.
    \item In most processes, the partial decay width for neutral pion emission is greater than that for charged pion emission.
    \item We identify several decay channels with sizable widths, such as {$P_{c1}^{\Sigma_c^*\bar{D}^*}[1/2(3/2^-)] \to P_{c}(4312)+\pi$, $P_{cs1}^{\Xi_c^{*}\bar{D}}[1(3/2^-)]\to P_{cs}(4338)+\pi$, $P_{cs0}^{\Xi_c^*\bar{D}^*}[0(1/2^-)] \to P_{cs1}^{\Xi_c^{\prime}\bar{D}}[1(1/2^-)]+\pi$ and $P_{cs1}^{\Xi_c^*\bar{D}^*}[0(3/2^-)] \to P_{cs1}^{\Xi_c^{\prime}\bar{D}}[1(1/2^-)]+\pi$}, which represent promising signatures for future experimental detection.
\end{itemize}

In summary, the systematic patterns we have identified in pion emission decays provide a new set of criteria for testing molecular interpretations and assigning quantum numbers to the observed (strange) hidden-charm pentaquarks. We strongly encourage experimental collaborations to test these predictions. Furthermore, this methodological approach can be extended to guide the search for other, yet-unobserved (strange) hidden-charm molecular pentaquarks, particularly those composed of excited constituents.

\section*{ACKNOWLEDGMENTS}

We would like to thank Hui-Hua Zhong and Xian-Hui Zhong for valuable discussions. This project is supported by the National Natural Science Foundation of China under Grants Nos. 12305139, 12175037, 12335001, and in part by National Key Research and Development Program under the contract No. 2024YFA1610503. Rui Chen is also supported by the Xiaoxiang Scholars Program of Hunan Normal University.



\end{document}